\documentclass[journal=nalefd,manuscript=communication]{achemso}

\usepackage[version=3]{mhchem} 
\usepackage{graphicx}
\usepackage{dcolumn}
\usepackage{multirow}
\usepackage{epsfig}
\usepackage{amssymb}
\usepackage{slashed}
\usepackage{braket}
\usepackage{amsmath}

\author{Ping Li}
\affiliation[Xiangtan]
{Department of Physics, Xiangtan University, Xiangtan, Hunan 411105, China}
\author{Xiao Li}
\affiliation[Austin]
{Department of  Physics, The University of Texas at Austin, Austin, Texas 7812}
\author{Wei Zhao}
\affiliation[Xiangtan]
{School of Mechanical Engineering, Xiangtan University, Xiangtan, Hunan 411105, China}
\author{Hua Chen}
\affiliation[Austin]
{Department of  Physics, The University of Texas at Austin, Austin, Texas 7812}
\author{M. X. Chen}
\affiliation[Hunan]
{College of Physics and Information Science,  Hunan Normal University, Changsha, 410081, China}
\author{Zhi-Xin Guo}
\email{zxguo08@hotmail.com}
\affiliation[xiangtan]
{Department of Physics, Xiangtan University, Xiangtan, Hunan 411105, China}
\author{Ji Feng}
\affiliation[beijing]
{International Center for Quantum Materials, School of Physics, Peking University, Beijing 100871, China}
\author{Xin-Gao Gong}
\affiliation[Fudan]
{Key Laboratory of Computational Physical Sciences and Department of Physics, Fudan University, Shanghai 200433, China}
\author{Allan H. MacDonald}
\affiliation[Austin]
{Department of  Physics, The University of Texas at Austin, Austin, Texas 7812}

\title[Dirac Electrons]
  {Topological Dirac states beyond $\pi$ orbitals for silicene on SiC(0001) surface}

\keywords{Silicene, Dirac electrons, topological properties, first-principles calculations} 

\begin{document}
\begin{abstract}
The discovery of intriguing properties related to the Dirac states in graphene has spurred huge interest in exploring its two-dimensional group-IV counterparts, such as silicene, germanene, and stanene. However, these materials have to be obtained via synthesizing on substrates with strong interfacial interactions, which usually destroy their intrinsic $\pi$($p_z$)-orbital Dirac states.
Here we report a theoretical study on the existence of Dirac states arising from the $p_{x,y}$ orbitals instead of $p_z$ orbitals in silicene on 4H-SiC(0001), which survive in spite of the strong interfacial interactions. We also show that the exchange field together with the spin-orbital coupling give rise to a detectable band gap of 1.3 meV. Berry curvature calculations demonstrate the nontrivial topological nature of such Dirac states with a Chern number $C = 2$, presenting the potential of realizing quantum anomalous Hall effect for silicene on SiC(0001). Finally, we construct a minimal effective model to capture the low-energy physics of this system. This finding is expected to be also applicable to germanene and stanene, and imply great application potentials in nanoelectronics. 
\end{abstract}

Graphene, a hexagonally bonded carbon atom sheet, exhibits fascinating properties due to its low-energy states behaving like massless Dirac electrons, and is also regarded as a promising material for the next-generation information technology due to its high mobility\cite{gr1,gr2,gr3,Qiao1}. Attention on hexagonal structures formed by the other group-IV elements Si, Ge, and Sn has been quickly increasing recently due to their potential easy integration in existing applications in the semiconductor industry \cite{Tao,Zhao,Ni}. Moreover, the stronger spin-orbital coupling (SOC) in the heavier Si, Ge, Sn may lead to the exotic quantum spin Hall phase\cite{Kane} 
that is experimentally detectable at ambient conditions, and further expand the frontier of Dirac-electron science\cite{Yao1,Yao2,Xu}.

It was predicted that freestanding Si, Ge, Sn monolayers which are called silicene, germanene, stanene, respectively, assume a buckled honeycomb lattice and possess the $\pi$($p_z$)-orbital Dirac states (denoted as $p_z$ Dirac states) similar to that in graphene\cite{Cahangirov,Yao1,Ezawa}. However, owing to the nature of preferring the $sp^3$ hybridization to the $sp^2$ hybridization\cite{Guo1,Guo2}, these two-dimensional (2D) materials are unable to be prepared by the mechanical exfoliation methods used for graphene. An alternative way is to synthesize them on substrates with strong interfacial interactions, since the van der Waals (vdW) interaction is usually too weak to stabilize their 2D honeycomb lattice\cite{zhao1}. Up to now, monolayer silicene, germanene, as well as stanene have been experimentally obtained on substrates such as Ag, ZrB$_2$, Ir, Au, Al, and Bi$_2$Te$_3$, etc. \cite{Vogt,Lin,Fleurence,Feng,Gao1,Gao2,Lay-ge,Der,Jia}. However, it is found that the strong hybridization between the $p_z$ orbitals in these materials and the substrates states make the desired $p_z$ Dirac states at the Fermi level ($E_F$) absent\cite{Fleurence,Guo3,Chen1,Guo4,Guo5,Lu,Molle}. 

Different from the $p_z$ orbitals which is easily hybridized with substrates, the $p_{x,y}$ orbitals forming the $\sigma$ bond in group-IV materials is particularly robust. It is thus ideal if one can obtain Dirac states formed primarily by $p_{x,y}$ orbitals (denoted as $p_{x,y}$ Dirac states) in group-IV 2D materials. Dirac states arising from $p_{x,y}$ orbitals have been predicted in cold atom systems with 2D honeycomb lattice\cite{Wu1, Wu2}, but have not been realized in solid state systems. 
Here we find that the $p_{x,y}$ Dirac states can exist in silicene grown on the C-terminated 4H-SiC(0001) [SiC(0001) hereandafter]\cite{sic1,sic2}. Moreover, we find that magnetism arising from unpaired $p$ electrons and the SOC give rise to a quantum anomalous Hall (QAH) gap of 1.3 meV with a Chern number $C = 2$, which should be experimentally detectable.

\begin{figure}
\begin{center}
\includegraphics[angle= 0,width=0.85\linewidth]{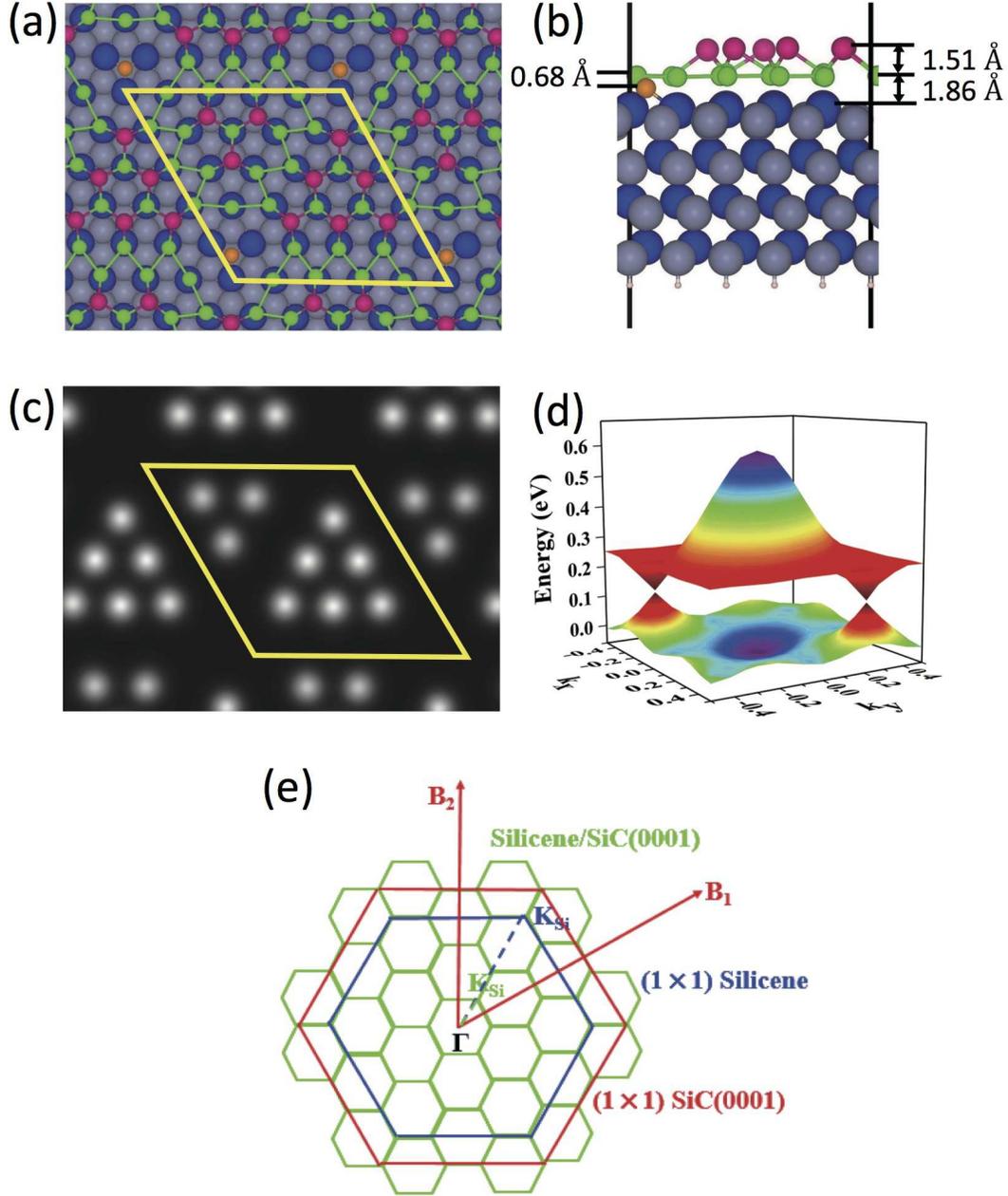}
\caption{
Relaxed atomic structures, simulated STM, calculated 3D band structure, and  schematic geometry of BZ for the c-top silicene/SiC(0001). (a), (b) Top and side views of the relaxed structure.  For the silicene, the protruded and dented Si atoms are depicted by the purple and orange balls, respectively, and the planer Si atoms in silicene are depicted by the green balls. For the SiC(0001), the C and Si atoms in SiC(0001) are depicted by the large blue and  grey balls, respectively, and the saturated hydrogen atoms on the bottom are depicted by the small light-pink balls.  (c) Simulated STM image. The lateral supercells in (a) and (c) are indicated by the yellow lines, and the vertical supercell in (b) is indicated by the black lines. (d)  3D band structure, where two pairs of Dirac cones at the $K$ and $K^{\prime}$ points are clearly seen around 0.1 eV above $E_F$.  (e) BZs of silicene/SiC(0001), $1\times1$ silicene, and $1\times1$ SiC(0001) shown in green, blue, and red, respectively. $B_1$ and $B_2$ represent the United vectors of $1\times1$ SiC(0001). }
\label{str}
\end{center}
\end{figure}

\textbf{Stable atomic structures of silicene/SiC(0001).} We have performed an extensive search for the stable structures of silicene/SiC(0001). We noticed that the $4\times 4$ unit cell of silicene is commensurate to the $5\times 5$ surface unit cell of SiC(0001) with a lattice mismatch of less than 1\%. We have considered 6 typical configurations, i.e., the top layer C/Si atom of SiC being in the top, bridge, and hollow positions of the silicene honeycomb lattice, respectively. Each supercell contains 257 atoms, including 32 silicene atoms, 200 SiC atoms, and 25 hydrogen atoms terminating the SiC substrate.  After extensive geometry optimizations, we have reached four distinct stable structures, which are denoted as c-top (Figure \ref {str}),  c-bridge (Figure S1), c-hollow (Figure S2), and si-bridge (Figure S3), respectively. A common feature in all the four structures is the significant structural reconstruction of silicene, where a large hollow ring composed of 10-14 Si atoms appears in the silicene sheet and 7-9 Si atoms protrude vertically from the sheet by 1.45-1.51 \AA. The calculated average bond lengths of silicene are 2.43-2.57 \AA, larger than that of pristine  (freestanding) silicene (Table 1). Table 1 also shows the distance between the average height of the silicene sheet and the top C layer of the SiC. The distances for all the four structures are in the range of 1.86-1.93 \AA, close to the sum of the covalent radii of C (0.77 \AA) and Si (1.18 \AA), indicating the covalent nature of the bonding interaction between silicene and the top layer of SiC(0001). 

\begin{table}
\caption{
Geometry symmetry, cohesive energy $E_{c}$ (eV/Si), binding energy $E_{b}$ (eV/Si),  and the structural parameters of the four structures for silicene on SiC(0001) and those for the pristine (PS) silicene. $d_{si-sic}$ is the average spacing between the planer Si layer of silicene and the topmost C layer of SiC surface.  $d_{si-si}$ is the average Si-Si bond length in silicene. The buckling (\AA) shows the Si coordinates in the surface-normal direction relative to that of planer Si layer\cite{note}.}
\begin{tabular}{cccccccc}
  \hline
  \hline
  &Symmetry   & $E_{c}$  &$E_{b}$  & $d_{si-sic} $  &$d_{si-si}$ & buckling   \\
  \hline
PS silicene &$D_{3d}$ & 4.76        & --        & --       &2.27     &0.49  \\
c-top         & $C_{v3}$      & 7.56       &2.80        & 1.86       &2.57        &-0.68, 1.51  \\
c-bridge    & P1      & 7.54        &2.78        & 1.91        &2.47        &-0.60, 1.45  \\
c-hollow    & P1      & 7.52       &2.76        & 1.89        &2.55        &-0.55, 1.50  \\
si-bridge   & P1      & 7.46       &2.73        & 1.93        &2.43        &1.45  \\
  \hline
  \hline
\end{tabular}
\end{table} 
 
To quantify the interaction between silicene and SiC(0001), we calculated the cohesive energy $E_c$, defined as $E_c = (E_{SiC} + N_{Si} \mu_{Si} - E_{tot}) / N_{Si}$, and the binding energy $E_b$, defined as $E_b = (E_{SiC} + E_{silicene} -E_{tot})/N_{Si}$, for the four structures identified above. Here $E_{tot}$, $E_{SiC(0001)}$, and $E_{silicene}$ are the total energies of the silicene/SiC(0001), the clean SiC(0001) surface, and the pristine silicene, respectively. $N_{Si}$ is the number of Si atoms in the silicene, and $\mu_{Si}$ is the chemical potential of Si which is chosen to be the total energy of an isolated Si atom. The cohesive energy defined above is the energy gain in forming silicene on the SiC surface from Si atoms. The binding energy measures the energy gain in putting the pristine silicene on the SiC surface. The calculated $E_c$ and $E_b$ are shown in Table 1. For all the four silicene/SiC structures, the cohesive energies are about 2.8 eV/Si and 2.0 eV/Si larger than that of pristine silicene and diamond Si (5.55 eV/Si), respectively, showing that SiC(0001) surface is a suitable platform for growing silicene. The unusually large $E_c$ and $E_b$ imply the preference of forming 2D layered structures over the cluster/diamond Si structures, when depositing Si atoms on SiC(0001) surface. The binding energies (2.73-2.80 eV/Si) are about 50 times larger than the binding energy of silicene on h-BN surface (56 meV/Si) \cite{Guo3}, confirming the covalent-bonding nature between silicene and SiC surface. 

The total energy calculations show that the c-top structure, on which we will focus in the following discussions, is the most stable with $E_c$ being 17-70 meV/Si larger than that of the other three metastable structures, meaning that it is likely to be the structure synthesized experimentally. 
To further show its thermodynamic stability during the growth, we have calculated the total energies of bilayer silicene on SiC(0001) (Figure S4). The thermodynamic stability is evaluated by $\partial^2 E /\partial L^2 \approx (E_0+E_2-2E_1)/L^2$, with $E_0$ the total energy of clean SiC(0001), $L$ the monolayer silicene thickness, and $E_1$ and $E_2$
the total energies of monolayer and bilayer silicene on SiC(0001), respectively. The calculated $E_0+E_2-2E_1$ for the c-top structure is about 1.25 eV per silicene atom, showing that  $ \partial^2 E/\partial L^2 > 0$, which verifies its thermodynamic stability\cite{zhangzhenyu}.

\begin{figure}
\begin{center}
\includegraphics[angle= 0,width=0.85\linewidth]{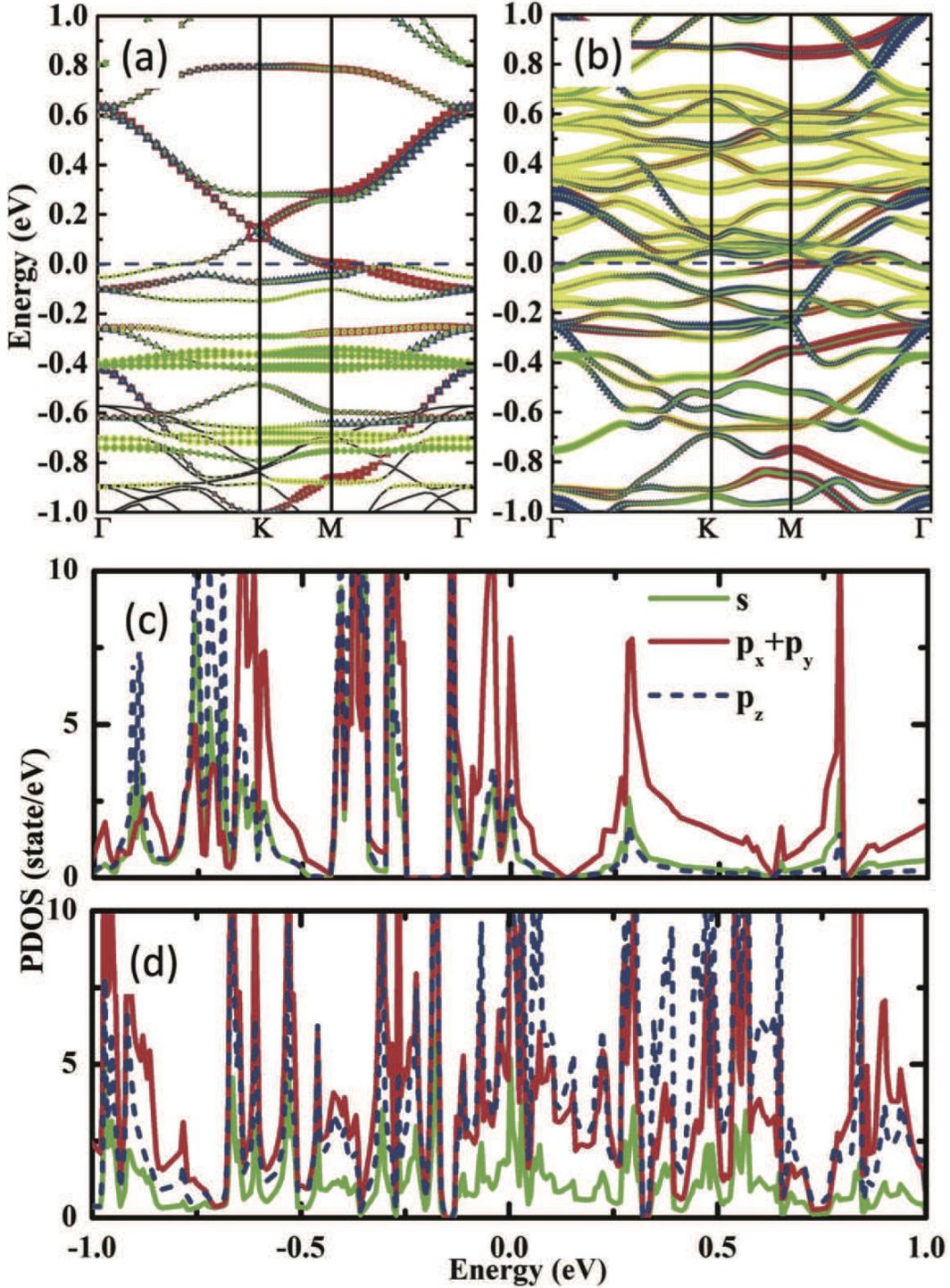}
\caption{
Calculated energy bands  and partial density of states (PDOS) projected on silicene for  the silicene/SiC(0001) and the isolated silicene of the c-top structure.  (a), (c) Energy bands and PDOS of  silicene/SiC(0001). (b), (d) Energy bands and PDOS of silicene detached from SiC(0001). $E_F=0$. The green rhombuses, red squares, blue triangles, and yellow circles in (a) and (b) indicate the contributions of $3s$, $3p_x$, $3p_y$, and $3p_z$ orbitals of silicene, respectively, the size of which are in proportion to the contributions. States marked by the red square at $K$ in (a) shows  characters of $3p_{x,y}$ orbitals. The color codes for orbitals in (d) are the same as those in (c).
}
\label{bd}
\end{center}
\end{figure}

\textbf{Low-energy Dirac electron states.} It is known that the strong interfacial interaction usually destroys the Dirac states of group-IV 2D materials \cite{Fleurence,Guo3,Chen1,Guo4,Guo5,Lu,Molle}. However, our calculated energy bands of the c-top structure show two pairs of Dirac cones around 0.1 eV above $E_F$ at $K$ and $K^{\prime}$ of the supercell Brillouin zone (BZ), with a band gap of 6 meV (Figure \ref {str}d). The Dirac cones can be adjusted to the $E_F$ via N-type doping in the SiC substrate (Figure S5), which is realized during its growth\cite{sic-dope}.  The energy bands around the high symmetry points in BZ are magnified in Figure \ref{bd}a, where two linear bands around $K$ point can be clearly seen nearby $E_F$. The BZs of silicene/SiC(0001), pristine silicene, and Si(0001) are shown in Figure \ref{str}e, where the $K$ point of pristine silicene lies on that of silicene/SiC. The electron velocity of such Dirac states is $1.56 \times 10^5$ m/s, comparable to that  in pristine silicene ($5\times 10^5$ m/s). 
The band structure is also desirable in that there are no other bands close to the Fermi energy. Thus the low-energy physics of the present system is fully governed by the Dirac states.  

\begin{figure}
\begin{center}
\includegraphics[angle= 0,width=0.95\linewidth]{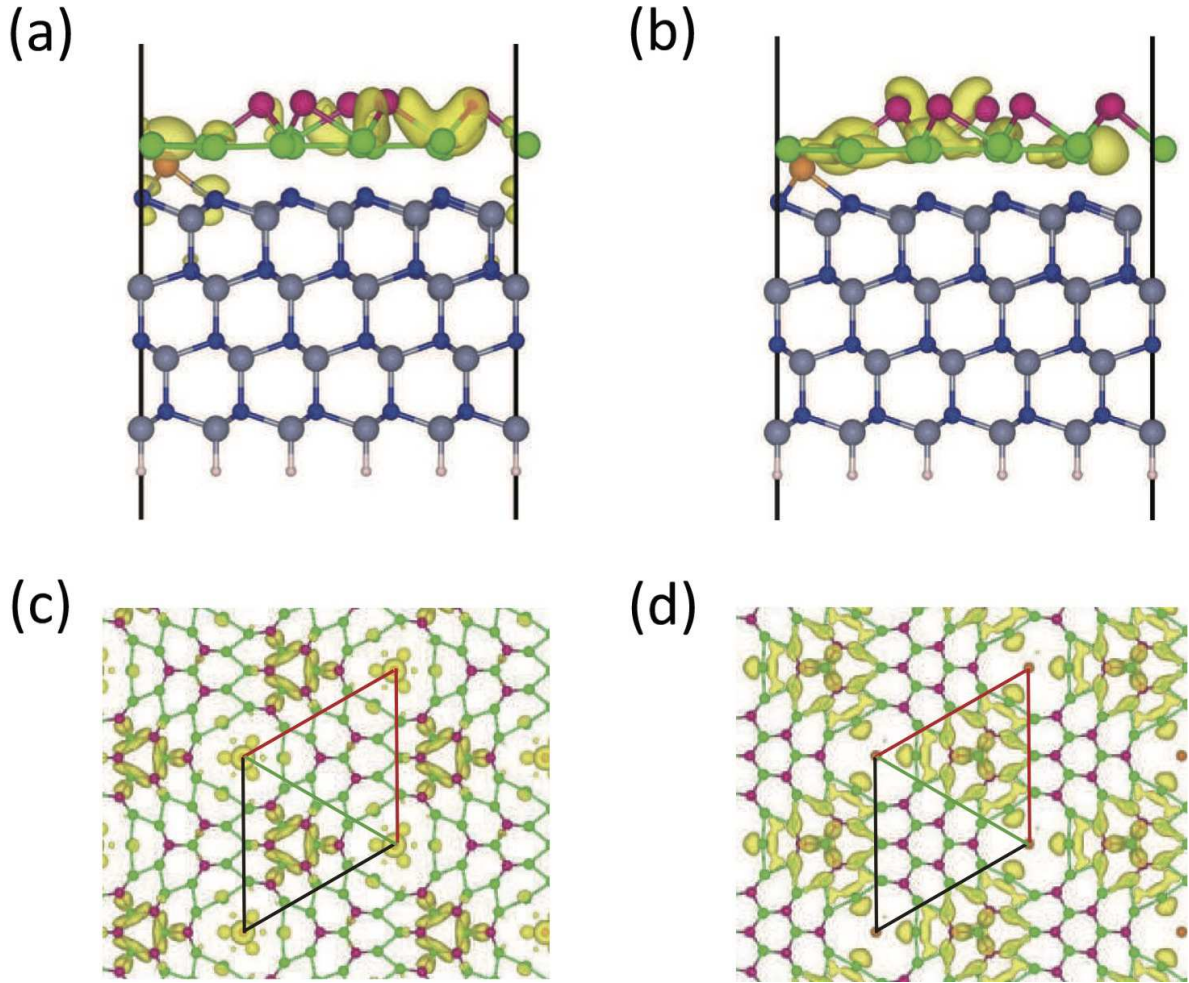}
\caption{
Kohn-Sham (KS) orbitals for the Dirac states in c-top silicene/SiC(0001) as marked by the red square in Figure \ref{bd}a. (a), (c) Side and top views of the KS orbitals for the lower energy level Dirac state at $K$ point. (b), (d) Side and top views of the KS orbitals for the higher energy level Dirac state at $K$ point. The isovalue surface level is at 0.0015 e$^{-1}$/\AA$^3$. The color codes of atomic balls are the same as in Figure \ref{str}. The KS orbitals mainly concentrate on the triangles indicated by the black and red lines in (c) and (d), respectively. Two triangles consist of a silicene supercell.}
\label{ch}
\end{center}
\end{figure}

To understand the atomic origin of the unusual Dirac states in silicene/SiC(0001), we have plotted the real-space distribution of the low-energy Kohn-Sham (KS) wavefunctions at the $K$ point. As shown in Figure \ref{ch}a,b, the KS wavefunctions of the Dirac states mainly locate on the silicene, suggesting their origin of atomic orbitals in silicene. We have further projected the KS wavefunctions of the Dirac states onto the Si $3s$, $3p_x$, $3p_y$, and $3p_z$ orbitals in silicene. From Figure \ref{bd}a, one can see that the Dirac states are mainly contributed by the $3p_x$ and $3p_y$ orbitals, while the energy bands mainly consisting of $3s$ and $3p_z$ orbitals are far from the Dirac cone. Especially, the $3p_z$-orbital characteristic bands in silicene appear in the energy range of [-0.9,-0.1] eV, and have lost the feature of Dirac states. We have additionally calculated the partial density of states (PDOS) for these orbitals projected on the silicene layer. The PDOS of 3$p_{x,y}$ orbitals is dominating in the energy range of [-0.1,0.6] eV, where a V-shape PDOS arising from 3$p_{x,y}$ orbitals clearly appears around the Dirac cone (Figure \ref{bd}c). These results show that the Dirac states in silicene on SiC(0001) are formed primarily by $p_{x,y}$ orbitals rather than the normally expected $p_z$ orbitals, in stark contrast to the case of pristine silicene, and suggesting the essential role played by the silicene/SiC(0001) interface. It is noteworthy that the present finding is different from that of silicene on the hydrogen-covered Si(111), where the $p_z$ Dirac states survive under the weak vdW interfacial interaction\cite{Guo3}.

The underlying mechanism for the formation of $p_{x,y}$ Dirac states is closely related to the strong hybridizations between the $s$, $p_z$ orbitals in the Si atoms of silicene and those in the C atoms of SiC surface, which significantly change the intrinsic hybridizations among the $3s$, $3p_{x,y}$, and $3p_z$ orbitals in silicene.  The $3p_z$ orbitals in silicene are pushed away from the Fermi energy due to hybridization with the SiC surface orbitals, while the $3p_{x,y}$ orbitals shift closer to the Fermi energy due to weakened $\sigma$ bonding within the silicene sheet.  This "orbital-filtering" effect is analogous to that appeared in DFT predictions of functionallized  atomic layers \cite{yao3} and heavy-metal atoms (Bi) on top of halogen-covered Si(111) surface\cite{liufeng}. 

As a confirmation of above mechanism, we further show in Figure \ref{bd}b the calculated energy bands of the silicene detached from SiC(0001). One can see that the 3$p_{x,y}$ Dirac bands have disappeared, and the states dominated by $3p_z$ orbitals emerge near the Fermi energy.  The disappearance of well-defined $p_z$ Dirac bands in the detached silicene is due to the large geometrical distortion which breaks the hexagonal symmetry of the $p_z$-characteristic orbitals\cite{Guo3}. The PDOS projection analysis shows that the energy bands near $E_F$ are contributed to the strong hybridizations among $3s$, 3$p_{x,y}$ and $3p_z$ orbitals of Si atoms in silicene  (Figure \ref{bd}d).
Above results confirm that the covalent-bonding interaction with the SiC substrate dramatically changes the low-energy states in silicene, pulling the $3s$ and $3p_z$ characteristic bands away from Dirac point around $E_F$ and creating the $3p_{x,y}$ Dirac states. 

It is noted that the band structure of silicene/SiC (Figure \ref{bd}a) is much simpler than that of the detached silicene (Figure \ref{bd}b) in the energy range of [-1.0, 1.0] eV.  Our DFT calculations show that the band gap of 4H-SiC(0001) is 2.4 eV, which covers the energy window shown in Figure \ref{bd}. From the orbital projection analysis, the energy bands in Figure \ref{bd}a mainly reflects the band structure of silicene. The strong orbital hybridizations between silicene and SiC induces the separation between $3p_{x,y}$-orbital characteristic Dirac bands and the $3s$-, $3p_z$-orbital characteristic bands, which makes the band structure simple in the energy window. 
Whereas, the complex band structure appears for the silicene detached from SiC (Figure \ref{bd}b), mainly due to the large geometrical distortion and the absence of SiC substrate, which push the previously saturated $3s$ and $3p$ orbitals of silicene closer to the Fermi energy.

A novel discovery of our study is the survive of $p_{x,y}$  Dirac states in a system with complex geometry, where the honeycomb structure of the pristine silicene is largely deformed. This character can be owning to two facts. On the one hand, the $p_{x,y}$ Dirac states are of particularly robust against the deformation due to the nature of $\sigma$-like orbitals hybridization, which is strong and hardly destroyed.  This feature is different from the $p_z$ Dirac states which are easily destroyed by the deformation and/or strong substrate interactions due to the nature of weak $\pi$ bonds\cite{Guo3}.  On the other hand, despite the large deformation exists, the silicene structure still preserves the $C_3$ rotational symmetry of pristine silicene, which is crucial for the appearance of Dirac bands as discussed in the later section. This is further supported by the absence of Dirac states in the c-bridge, c-hollow, and si-top structures, all of which have the P1 symmetry (Figures S1-S3).

\begin{figure}
\begin{center}
\includegraphics[angle= 0,width=0.90\linewidth]{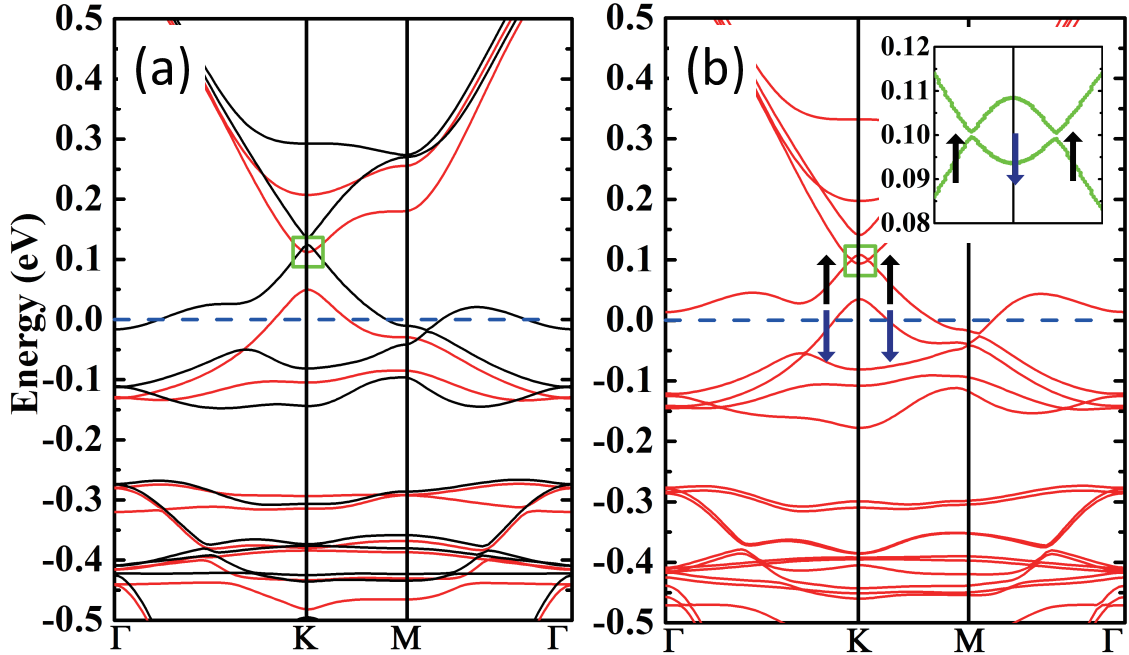}
\caption{
Calculated energy bands of c-top silicene/SiC(0001) with and without SOC. (a) Spin-polarized energy bands without SOC for silicene on the 4-layer SiC(0001). The black and red curves correspond to the spin-up and spin-down energy bands, respectively. The spin-up and spin-down Dirac states overlap at $K$ point. The green square shows a crossover between spin-up  and spin-down Dirac states. (b) Energy bands with SOC for silicene on the 1LSiC(0001). The inset in (b) is a zoom-in of the band structure in the green square. The spin polarizations of the lower-energy Dirac states are indicated by the arrows. $E_F=0$.
}
\label{bd2}
\end{center}
\end{figure}

\textbf{Quantum anomalous Hall gap in silicene/SiC(0001).} A geometrical feature of the c-top structure is that one Si atom in the center of the large Si ring in the silicene sheet dents towards the substrate and bonds to three underlying C atoms of the SiC, forming isolated Si-C pairs (Figure \ref{str}). The special bonding environment can lead to unpaired $p$ electrons which may favor a magnetic ground state over the nonmagnetic state. Indeed, our calculations show that the ferromagnetic state, with magnetic moment about 1 $\mu B$ per supercell cell, is more stable than the nonmagnetic state by 1 meV per Si atom in silicene. The magnetization  is induced by the dangling bond of the dented Si atom of silicene, neighboring three C atoms in the substrate. The dangling-bond state shows a very strong exchange splitting and contributes to the total magnetic moment\cite{mag}. 
Magnetism is not observed in the other three silicene/SiC(0001) structures where no isolated Si-C pairs exist. 

The exchange field arising from the nonzero magnetic moment further splits the Dirac bands near $E_F$ into spin-up and spin-down bands, as shown in Figure \ref{bd2}a from spin-polarized calculations (without SOC). Note that the spin-up band for the lower-energy Dirac states and the spin-down band for the higher-energy Dirac states cross near the Fermi energy, which as we show below is important for the establishment of the QAH gap in the presence of spin-orbit coupling \cite{ezawa2,Qiao}. 

\begin{figure}
\begin{center}
\includegraphics[angle= 0,width=0.90\linewidth]{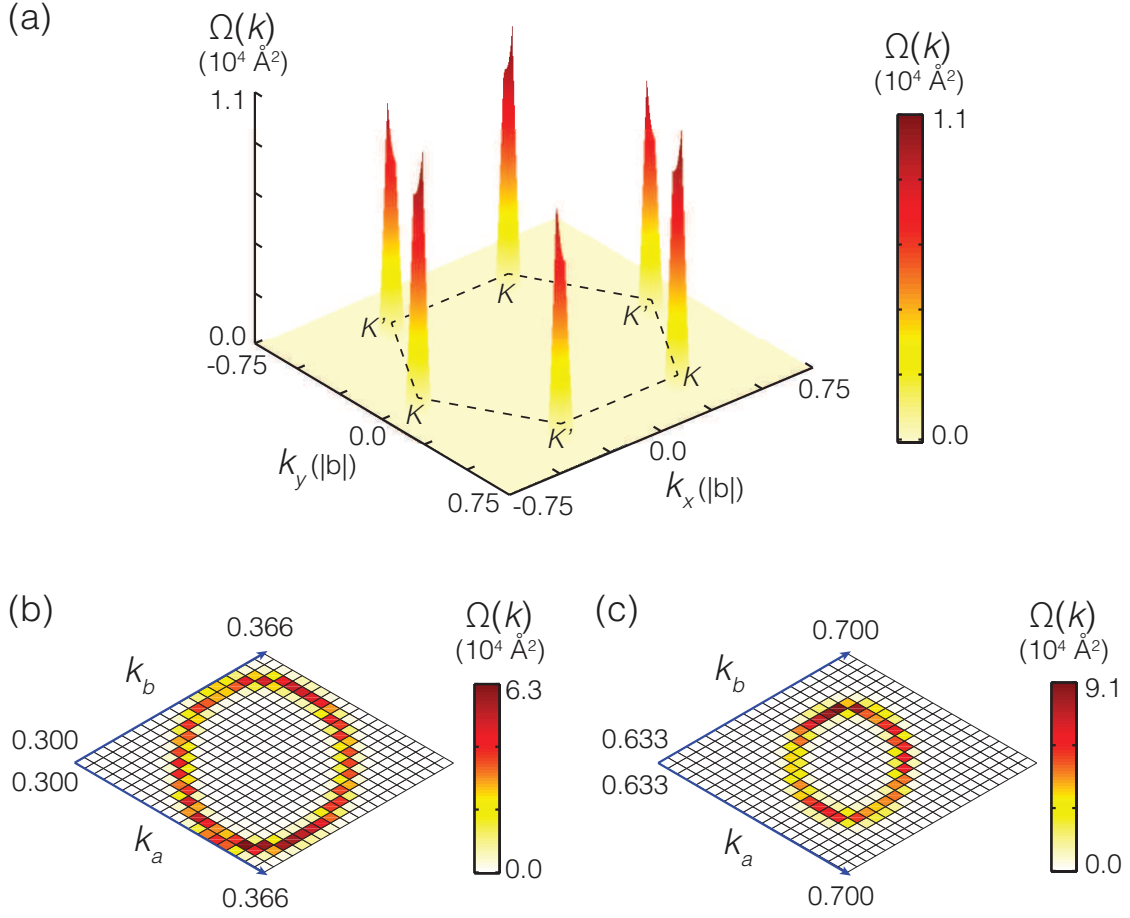}
\caption{
Calculated non-Abelian Berry curvature of the c-top silicene/SiC(0001) for energy bands below the Dirac SOC gap  over the entire BZ. (a) The 3D distribution of Berry curvature in momentum space. The first BZ is outlined by the dashed lines.
The Berry curvature is sharply concentrated around the Dirac cones on $K$ and $K^{\prime}$ points.
(b),(c) The zoom-in of 2D distribution of Berry curvature around the $K$ and $K^{\prime}$ points, respectively. 
}
\label{tp}
\end{center}
\end{figure}

It is known that the SOC in pristine silicene opens a band gap of 1.55 meV in the $p_z$ Dirac bands, and makes it a time-reversal invariant $Z_2$ topological insulator \cite{Yao2}. One may wonder what would happen for the $p_{x,y}$ Dirac states in silicene on SiC(0001) where the time-reversal symmetry has been broken by the ferromagnetism. We thus performed calculations with SOC and investigated the topological properties of the $p_{x,y}$ Dirac states. In Figure \ref{bd2}b we show the calculated energy bands of silicene on SiC with SOC (see \textbf{Methods} section). It can be seen that the SOC opens a band gap of 1.3 meV around $K$ point between the spin-up and the spin-down Dirac states, which is comparable to the $Z_2$ gap of the $p_z$ Dirac states in pristine silicene. 
Note that a band gap between $s_z$ eigenstates has to be opened by SOC terms involving $s_{x}$ and/or $s_y$ components, which are allowed by the broken mirror symmetry at the silicene/SiC(0001) interface and can be qualitatively captured by a Rashba-like term\cite{ezawa2,Qiao}. 

We next show that the gap at the Dirac cone of c-top structure is a QAH gap, characterized by a nonzero Chern number $C=2$. To this end we calculate the $U(N)$ Berry curvature of KS Bloch states, $\Omega(\textbf{k})$ \cite{Van,fengji}
\begin{equation}
\Omega(\textbf{k})\delta k_a \delta k_b=Arg\{det [L_a(\textbf{k})L_b(\textbf{k}+\delta \hat{k}_a)L^\dagger_a(\textbf{k}+\delta \hat{k}_b)L^\dagger_b(\textbf{k})]\},
\end{equation}
where $L_{\alpha}(\textbf{k})=\vec{u}(\textbf{k})\cdot \vec{u}(\textbf{k}+\delta \hat{k}_{\alpha})$ with $\alpha=a$ or $b$, $\delta \hat{k}_\alpha =\delta {k}_{\alpha} \hat{e}_{\alpha}$, and $\vec{u}(\textbf{k})=[\ket{u_1(\textbf{k})},...,\ket{u_N(\textbf{k})}]$ which is the cell-periodic part of the Bloch wave function of $j$th band at $\textbf{k}$. A quadrant of BZ is pixelated into $30\times 30$ plaquettes for the evaluation of $U(N)$ Berry curvature. A total of 354 bands up to the band gap at the Dirac cone (0.1 eV) are considered, which cover all the valance states from the outer shell orbitals of carbon and silicon atoms.  
Figure \ref{tp} shows the Berry curvature distribution in the momentum space, which is peaked at the corners of the first BZ around $K$ and $K^{\prime}$ points.
The Chern number $C$ is calculated by integrating the non-Abelian Berry curvature by\cite{Niu}:
\begin{equation}
C= \frac{1}{2\pi} \int_{BZ} d^2k ~\Omega(\textbf{k})
\end {equation}
The absolute value of $C$ corresponds to the number of gapless chiral edge states along an edge of the silicene sheet. We found $C=2$, meaning that there are two chiral edge states.   
In general each band of a gapped $2\times2$ Dirac model has a Chern number equal to 1/2 or -1/2. However, in the present case the low-energy space in each valley consists of four states involving both spin and pseudospin degrees of freedom (as discussed in the following section), and the above argument is not directly applicable. 
To see how this Chern number comes about, we note that in the absence of a gap term in the pseudospin space, the lower two bands of a $4\times4$ Dirac Hamiltonian with a perpendicular exchange field and the Rashba SOC contribute to a Chern number of 1 per valley, or total $C=2$ \cite{Qiao,ezawa2}. Here we only argue that the pseudospin gap term in the low-energy theory of our system does not lead to further change of the Chern number\cite{chern}.

As expected from the nonzero Chern number, the anomalous Hall conductivity should also show a quantized charge Hall plateau of $\sigma_{xy} = Ce^2/h$ when the $E_F$ is located in the QAH gap. Figure \ref {tp}  shows that the main contribution to the Berry curvature is sharply concentrated around the Dirac cones, indicating that the main source of the anomalous Hall conductivity arises from states near the SOC gaps. Compared to previous proposals of realizing the QAH effect, the present system has the advantage of no additional magnetic doping.

It was found that the band structure can be controlled by applying an electric field ($E$) perpendicular to the pristine silicene\cite{ezawa3,ezawa4,dum}. Especially,  Ezawa found that in the pristine silicene, where the Dirac band gap is dominated by the effective SOC,  the band gap decreases to zero with $E$ increasing to a critical value ($E_c=$ 0.017 V/\AA), and then increases with $E$ for $E>E_c$\cite{ezawa3,ezawa4}. To explore the electric field effect on the Dirac band gap in the more complex structure, we have performed calculations by applying the perpendicular electric fields E of different strength ($E$= 0.005, 0.01,  0.02,  0.05, and 0.1 V/\AA). The results show that the electric field has minor effects on the topological Dirac gap (varying within 0.1 meV) and does not induce gap closing (Figure S7), which is different from the case of pristine silicene.  There are two possible reasons for this phenomenon. One is that in our case the gap is dominated by the Rashba SOC and the exchange coupling which are expected to be insensitive to the electric field\cite{ezawa3}, the other is that our system is metallic which can largely screen the electric field.

\begin{figure}
\begin{center}
\includegraphics[angle=-90,width=0.90\linewidth]{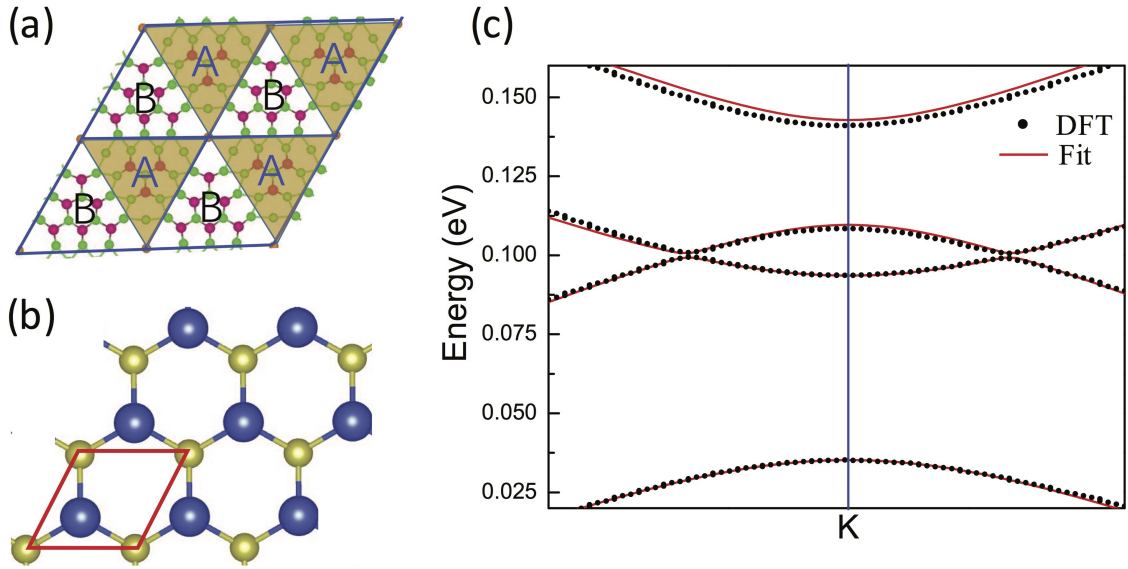}
\caption{
Schematic geometry of the c-top silicene structure and fitted band structure. (a) Schematic geometry of the c-top silicene which can be divided into two triangle parts, as marked by A and B, respectively. The supercell of silicene is indicated by the blue quadrilateral. (b) Schematic geometry of the 2D lattice composed of A (big blue balls) and B (small yellow balls), which consist a honeycomb structure. The unit cell is indicated by the red quadrilateral. (c) Fitted band structure around Dirac cone at $K$ in use of the low-energy effective Hamiltonian, in comparison with the DFT results  for the c-top silicene/SiC(0001).  
}
\label{tb}
\end{center}
\end{figure}

\textbf{Low-energy effective model.} It is desirable to come up with a low-energy effective model to understand the origin of the Dirac states and the QAH effect. However, owing to that present system consists of too many atoms, the standard procedure which constructs the tight-binding Hamiltonian from KS eigenstates and thus extracts the effective Hamiltonian based on the $k\cdot p$ approach, is not very practical.
We thus propose an alternate strategy. Because the Dirac states in our system are well separated from the other bands at $K$ and $K^{\prime}$, the low-energy $k\cdot p$ Hamiltonian has the form of a $4\times4$ Hermitian matrix. Because the SOC in our system is expected to be much weaker than the other energy scales, the 4 basis functions should be well approximated by direct products of orbital and spin parts. The 2D orbital part can be viewed as a 1/2 pseudospin. In the weak SOC limit, all terms in the Hamiltonian can be written as direct products of Pauli matrices and identity matrices in the pseudospin and the real spin spaces.

In Figure \ref{ch}c,d we have shown the real space distributions of the two pseudospin eigenfunctions. One can see that they nicely fall into the two neighboring equilateral triangles that compose the supercell with $C_3$ rotational symmetry. This is very similar to the case of graphene with a staggered sublattice potential. The difference is just that in graphene the pseudospin basis functions are the $p_z$ orbitals in the two sublattices of the carbon honeycomb lattice.  In the present case, the two basis functions are certain linear combinations of (mainly) Si $p_{x,y}$ orbitals in the two equilateral triangles of the unit cell, which also effectively form a much bigger honeycomb lattice (Figure \ref{tb}a,b). 
Consequently, one can get the low-energy effective Hamiltonian for the Dirac states around the $K_\eta$ point  in the pseudospin block (see Supporting Information), as
\begin{equation}
H_\eta=\epsilon^+ I+\epsilon^-\tau_z+M^+\sigma_z+M^-\tau_z\sigma_z+\lambda_R(\eta\tau_x\sigma_y-\tau_y\sigma_x)/2+\lambda_{SO}\eta \tau_z\sigma_z+\hbar v_F(\eta k_x\tau_x+k_y\tau_y)
\end{equation}
where  $\boldsymbol{\sigma}$=($\sigma_x$, $\sigma_y$, $\sigma_z$) is the Pauli matrix of spin, $\tau_a$ ($a=x,y,z$) is the Pauli matrix of the sublattice pseudospin, $\eta=\pm1$ labels the two valleys at $K$ and $K^{\prime}$, and $v_F=\sqrt3a_0t/2\hbar$ with  $a_0$ the lattice constant of the supercell.
 In the low-energy effective Hamiltonian, the parameters are set as $t=1.5$ eV for the transfer energy,  $\epsilon^\pm=(\epsilon^A\pm\epsilon^B)/2$ with $\epsilon^A=118.2$ meV and  $\epsilon^B=72.4$ meV for the on-site energy, $M^\pm=(M^A\pm M^B)/2$ with $M^A=24.6$ meV and $M^B=37.2$ meV for the exchange energy, $\lambda_{R}=0.7$ meV and $\lambda_{SO}=5.5$ meV for the Rashba and  intrinsic SOC, respectively.  As shown in Figure \ref{tb}c, the band structure around the Dirac cone has been well reproduced in use of these parameters.
Compared with the low-energy effective Hamiltonian for the X-hydride/halide system\cite{yao3}, the Hamiltonian for the present system has additional terms from the staggered field, exchange field, and Rashba SOC.
 
A four-band second-nearest-neighbor tight-binding model for the $p_{x,y}$-orbital basis  can be also constructed as
\begin{equation}
\begin{split}
H=\sum_{i\alpha}c_{i\alpha}^\dagger \epsilon_ic_{i\alpha}-t\sum_{\left \langle i,j \right \rangle\alpha}c_{i\alpha}^\dagger c_{j\alpha}+\sum_{i\alpha}c_{i\alpha}^\dagger M_{i}\sigma_z c_{i\alpha}\\
+i\lambda_R\sum_{\left \langle i,j\right \rangle \alpha\beta}c_{i\alpha}^\dagger (\boldsymbol{\sigma} \times \boldsymbol{d_{ij}})^z_{\alpha\beta}c_{j\beta}+\frac{i\lambda_{SO}}{3\sqrt{3}} \sum_{\left \langle \left \langle i,j \right \rangle \right \rangle\alpha\beta}c_{i\alpha}^\dagger \nu_{ij}\sigma^z_{\alpha\beta} c_{j\beta}
\end{split}
\end{equation}
where $\epsilon_i$=$\epsilon^A$ ($\epsilon^B$), $M_i$= $M^A$ ($M^B$) for A (B) sublattice, $c_{i\alpha}^\dagger$ and $c_{i\alpha}$ are creation and annihilation operators with spin polarization $\alpha$ on site  $i$, and the combinations $\left \langle i,j \right \rangle$  and $\left \langle \left \langle i,j\right \rangle \right \rangle$ run over all the nearest and next-nearest neighbor hopping sites, respectively. In the  Hamiltonian, the first term represents the on-site energy induced by staggered sublattice potential and substrate deformation potential.  The second term is the usual nearest-neighbor hopping term. The third term represents the exchange magnetization. The forth term is the Rashba SOC with $\boldsymbol{d_{ij}}$ representing a unit vector pointing from site B to site A. The fifth term represents the intrinsic SOC,  and $\nu_{ij}=\pm1$ if the next nearest neighboring hopping is anticlockwise (clockwise) with respect to the positive $z$ axis.

In summary, we have performed systematic electronic structure calculations on the basis of the density-functional theory for the system of silicene on C-terminated SiC(0001) surface. We found that the strong interfacial bonding stabilizes the silicene overlayer and results in Dirac states at $K$ point, which are composed of $p_{x,y}$ orbitals, rather than $p_z$ orbitals, of the Si atoms in silicene. We also found that the exchange field due to the unpaired electrons at the interface, together with the spin-orbital coupling give rise to a detectable band gap of 1.3 meV at the Dirac points. Berry curvature calculations using the Kohn-Sham wavefunctions further show the nontrivial topological nature of such Dirac states with a Chern number $C = 2$, indicating the silicene/SiC(0001) system is a QAH insulator without additional magnetic doping. We further constructed a minimal effective model to capture the low-energy physics of this system.  We expect the underlying physical mechanism to be also applicable to other group-IV 2D materials such as germanene and stanene, which will be a interesting subject for future studies.

\textbf{Methods.} \emph{nonSOC calculations.} In our work, density functional theory (DFT) method was used for structural relaxation and electronic structure calculation. The ion-electron interaction was treated by the projector augmented-wave technique as implemented in the Vienna ab initio simulation package (VASP)\cite{vp1,vp2}.  The vdW density functional (optB86b-vdW functional) \cite{Dion, Klimes} which is capable of treating the dispersion force, was adopted for the exchange-correlation functional. The electron-ion interaction was described by the projector augmented wave method\cite{Perdew}, and a cutoff energy of 450 eV in the plane-wave basis set was used. The Brillouin zone was sampled with an $3\times 3\times 1$ Monkhorst-Pack k-mesh in the geometry optimization calculations. All atomic positions and lattice constants were optimized by using the conjugate gradient method where the total energy and atomic forces are minimized. The geometry optimization performed until the remaining Hellmann-Feynman forces become less than 0.01 eV/\AA~to obtain the final structures. The electronic structures were calculated using $6\times 6\times 1$ Monkhorst-Pack  k-mesh in Brillouin zone.
The  SiC surface was simulated by a repeating slab model consisting of a four-SiC-layer slab cleaved from the C-terminated 4H-SiC(0001) with the calculated lattice constant of 3.078 \AA~using vdW-DF functional. The silicon atoms on the bottom layer were saturated by hydrogen atoms. The SiC slab was separated from its images by the 20-25 \AA~vacuum region with monolayer to trilayer Si atoms placed on top.  The silicene structures with $4\times 4$ periodicity were deposited on the $5\times 5$ SiC(0001) surface in the simulation. 

\emph{SOC calculations.} In the nonSOC calculations, the substrate was modeled by four SiC layers with a $5\times 5$ periodicity, which is too large for us to perform the SOC calculations. To overcome this computational issue, we constructed a thinner substrate by peeling the silicene and the top-layer SiC from the c-top structure and saturating the peeled SiC layer with hydrogen (denoted as silicene/1LSiC, Figure S6). We also performed nonSOC spin-polarized calculations for silicene/1LSiC (Figure S6c). A comparison of Figure S6c and Figure \ref{bd2}a shows that the Dirac states remain almost unchanged when a thinner slab is used for the substrate.

\emph{Low-energy effective model.}
Details are presented in Supporting Information.

\subsection{AUTHOR INFORMATION}
\textbf{Corresponding Authors}\\
zxguo08@hotmail.com
\\
\textbf{Author Contributions}\\
P. L., X. L.  and W. Z. contributed equally to this work. 
\\
\textbf{Present Address}\\
 P. L. is now in Department of Physics, Soochow University
 
\begin{acknowledgement}
We are grateful for useful discussions with Prof. Qian Niu, Prof. Erio Tosatt, and Mr. Wei Luo. 
This work is surportted by National Natural Science Foundation of China (Grant No. 11604278, 11374252).
\end{acknowledgement}

\begin{suppinfo}
Figures displaying geometrical and electronic structures, and simulated STM  images of c-bridge, c-hollow, si-bridge silicene/SiC(0001) structures; Atomic structure and band structure of silicene on doped SiC(0001); Additional information on the total energy for the multilayer silicene on SiC(0001);  Geometrical and electronic structures of silicene/1LSiC; Dirac gap variation with the applied electric field of silicene/1LSiC, and Methods on low-energy effective Hamiltonian.
\end{suppinfo}


\begin{thebibliography}{}
\bibitem{note}
The results in Table 1 are from non-spin polarization calculations. We have also performed spin polarization calculations for the four silicene/SiC(0001) structures. It is found that only the c-top structure has a stable ferromagnetic phase, while its energy and geometry results are very close to those in the non-ferromagnetic phase listed in the table.
\bibitem{chern}
The argument is based on the fact that the Chern number can only change upon closing and reopening of the gap between the filled and the empty bands, because of its topological nature. However, without the Rashba SOC, a gap term in the pseudospin space by itself cannot open a gap perturbatively between the 2nd and the 3rd bands in the Zeeman-splitted $4\times4$ Dirac Hamiltonian. This is because the ring-degeneracy between these two bands is between eigenstates of spin along z, which commutes with the pseudospin gap term. Therefore in the perturbative regime, i.e., the size of the pseudospin gap is much smaller than the exchange field, it cannot compete with the Rashba term in closing and reopening the gap between the 2nd and the 3rd bands, and cannot change the total Chern number of the filled (i.e. the 1st and the 2nd) bands. Thus the total Chern number remains to be 2. Figure 4a indicates that the system considered in this work is indeed in this regime, since the ring-degeneracy between Dirac bands with opposite spin polarizations stays intact in the absence of spin-orbit coupling. We do note that the pseudospin gap can compete with the Rashba term in closing and reopening the gap between the 1st and the 2nd bands and change their Chern numbers individually, but not their sum. We did not consider the intrinsic SOC term in this discussion since it also commutes with the z-spin operator.
\end{thebibliography}

\end{document}